\documentclass[10pt,twocolumn,letterpaper]{article}

\usepackage{cvpr}              
\usepackage{tikz}

\usepackage{etoolbox}

\makeatletter

\newcommand{\define}[4][ignore]{%
  \ifstrequal{#1}{ignore}{}{
  \@namedef{thmtitle@#2}{#1}}%
  \@namedef{thm@#2}{#4}%
  \@namedef{thmtypen@#2}{lemma}%
  \newtheorem{thmtype@#2}[theorem]{#3}%
  \newtheorem*{thmtypealt@#2}{#3~\ref{#2}}%
}

\newcommand{\state}[1]{%
  \@namedef{curthm}{#1}
  \@ifundefined{thmtitle@#1}{
  \begin{thmtype@#1}
    }{
  \begin{thmtype@#1}[\@nameuse{thmtitle@#1}]
  }
    \label{#1}
    \@nameuse{thm@#1}
  \end{thmtype@#1}
  \@ifundefined{thmdone@#1}{
  \@namedef{thmdone@#1}{stated}%
  }{}
}

\newcommand{\restate}[1]{%
  \@namedef{curthm}{#1}
  \@ifundefined{thmtitle@#1}{
    \begin{thmtypealt@#1}
    }{
  \begin{thmtypealt@#1}[\@nameuse{thmtitle@#1}]
  }
    \@nameuse{thm@#1}
  \end{thmtypealt@#1}
  \@ifundefined{thmdone@#1}{
  \@namedef{thmdone@#1}{stated}%
  }{}
}

\newcommand{\thmlabel}[1]{
  \@ifundefined{thmdone@\@nameuse{curthm}}{\label{#1}
    }{\tag*{\eqref{#1}}}
}
\makeatother

\definecolor{cvprblue}{rgb}{0.21,0.49,0.74}
\usepackage[pagebackref,breaklinks,colorlinks,allcolors=cvprblue]{hyperref}

\newcommand{\E}{\mathbb{E}}

\def\paperID{17173} 
\def\confName{CVPR}
\def\confYear{2025}

\title{Double Blind Imaging with Generative Modeling}

\author{Brett Levac\\
University of Texas at Austin\\
2501 Speedway, Austin, TX 78712, USA\\
{\tt\small blevac@utexas.edu}
\and
Ajil Jalal\\
University of California Berkeley\\
Berkeley, CA 94720, USA\\
{\tt\small ajiljalal@berkeley.edu}
\and
Kannan Ramchandran\\
University of California Berkeley\\
Berkeley, CA 94720, USA\\
{\tt\small kannanr@eecs.berkeley.edu}
\and
Jonathan I.\ Tamir\\
University of Texas at Austin\\
2501 Speedway, Austin, TX 78712, USA\\
{\tt\small jtamir@utexas.edu}
}

\begin{document}
\maketitle
\begin{abstract}

Blind inverse problems in imaging arise from uncertainties in the system used to collect (noisy) measurements of images. Recovering clean images from these measurements typically requires identifying the imaging system, either implicitly or explicitly. A common solution leverages generative models as priors for both the images and the imaging system parameters (e.g., a class of point spread functions). To learn these priors in a straightforward manner requires access to a dataset of clean images as well as samples of the imaging system.
We propose an AmbientGAN-based generative technique to identify the distribution of parameters in unknown imaging systems, using only unpaired clean images and corrupted measurements. This learned distribution can then be used in model-based recovery algorithms to solve blind inverse problems such as blind deconvolution. We successfully demonstrate our technique for learning Gaussian blur and motion blur priors from noisy measurements and show their utility in solving blind deconvolution with diffusion posterior sampling.

\end{abstract}    
\section{Introduction}
\label{sec:intro}

Computational imaging techniques are integral across a wide range of disciplines, including astronomy \cite{EventHorizon2019,Vojtekova_2020}, microscopy \cite{Rivenson2017, Chowdhury2019}, medicine \cite{PET2021, heckel2024deeplearningacceleratedrobust}, and consumer electronics \cite{delbracio2021}. Under additive noise, the imaging system can be modeled as
\begin{equation}
    y = A(x) + \eta,
\label{eqn:forward_process}
\end{equation}
where $x \in \mathbb{C}^n$ is the image, $A: \mathbb{C}^n \rightarrow \mathbb{C}^m$ represents the measurement (forward) operator, $\eta \in \mathbb{C}^m$ is random additive noise, and $y \in \mathbb{C}^m$ are the measurements.
The central challenge is to estimate the clean image $x$ from measurements $y$, a task known as the inverse problem. This problem is typically ill-posed due to additive noise ($\eta$) and potentially insufficient measurements ($m<n$) for stable inversion.

Numerous classical \cite{donoho2006, lustig2007} and deep learning-based methods \cite{Gregor2010, bora2017, kawar2022, Ongie2020} have been proposed to address signal recovery. These methods commonly assume that the measurement process $A$ is known a priori, allowing it to be used directly in the recovery algorithm. This assumption implies that the exact characteristics of the imaging system used to acquire the measurements are known. 
\begin{figure}[h]
    \centering
\includegraphics[width=0.99\columnwidth]{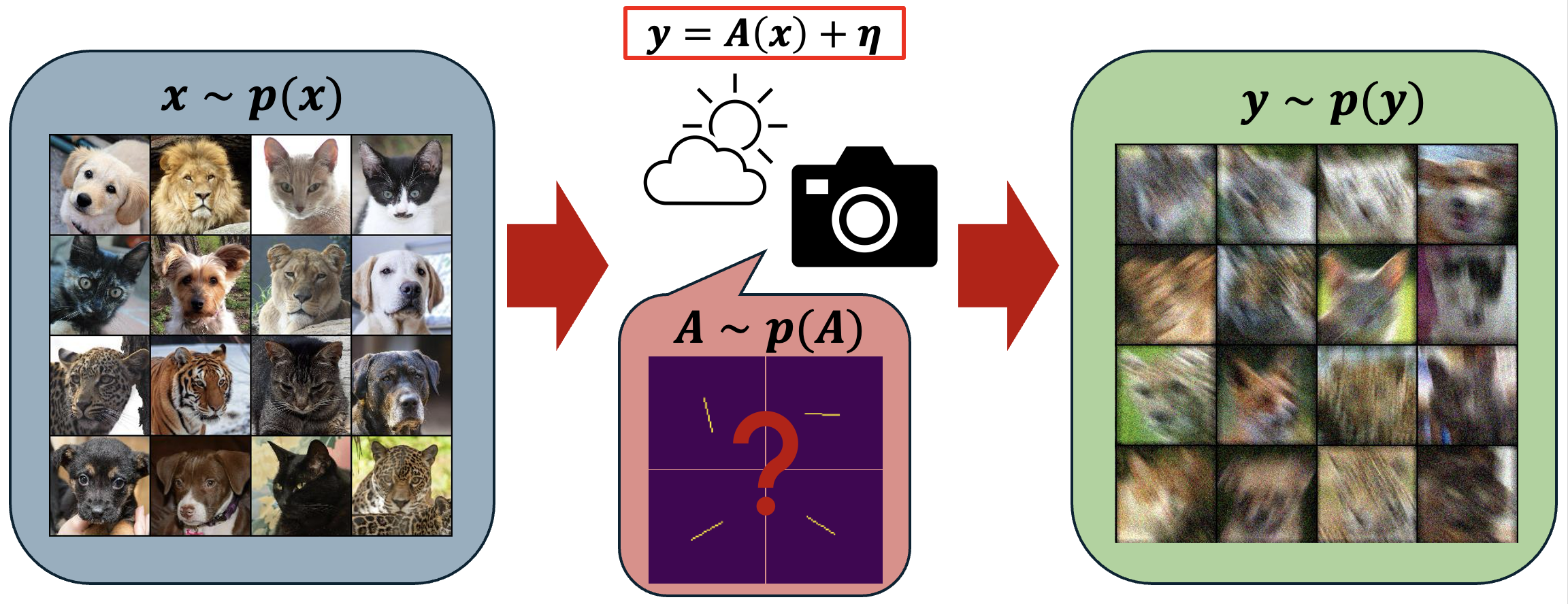}
    \caption{Example of imaging system setup where the measurement process is unknown due to hardware, environment, or scene variables.}
    \label{fig:proposed_setup}
\end{figure}

\begin{figure*}[htb]
\centering
\resizebox{0.7\textwidth}{!}{\tikzset{every picture/.style={line width=0.75pt}} 

\begin{tikzpicture}[x=0.75pt,y=0.75pt,yscale=-1,xscale=1]

\draw  [fill={rgb, 255:red, 74; green, 144; blue, 226 }  ,fill opacity=1 ] (960.5,205.97) -- (827.34,281.27) -- (827.71,130.34) -- cycle ;
\draw  [fill={rgb, 255:red, 224; green, 100; blue, 115 }  ,fill opacity=1 ] (184.57,68.73) .. controls (186.62,68.73) and (188.29,70.4) .. (188.29,72.45) -- (188.29,98.4) .. controls (188.29,100.46) and (186.62,102.13) .. (184.57,102.13) -- (173.39,102.13) .. controls (171.33,102.13) and (169.67,100.46) .. (169.67,98.4) -- (169.67,72.45) .. controls (169.67,70.4) and (171.33,68.73) .. (173.39,68.73) -- cycle ;
\draw  [fill={rgb, 255:red, 224; green, 100; blue, 115 }  ,fill opacity=1 ] (382.29,126.03) -- (220.78,99.92) -- (221.27,66.73) -- (383.53,43.05) -- cycle ;
\draw  [fill={rgb, 255:red, 184; green, 233; blue, 134 }  ,fill opacity=1 ] (441.62,170.95) .. controls (444.94,170.95) and (447.64,173.65) .. (447.64,176.97) -- (447.64,313.72) .. controls (447.64,317.04) and (444.94,319.73) .. (441.62,319.73) -- (423.57,319.73) .. controls (420.24,319.73) and (417.55,317.04) .. (417.55,313.72) -- (417.55,176.97) .. controls (417.55,173.65) and (420.24,170.95) .. (423.57,170.95) -- cycle ;
\draw  [fill={rgb, 255:red, 235; green, 221; blue, 70 }  ,fill opacity=1 ] (523.5,92.14) -- (630.51,125.28) -- (629.28,199.65) -- (521.19,231.03) -- cycle ;
\draw  [fill={rgb, 255:red, 224; green, 100; blue, 115 }  ,fill opacity=1 ] (440.19,44.44) .. controls (443.51,44.44) and (446.21,47.13) .. (446.21,50.45) -- (446.21,122.42) .. controls (446.21,125.75) and (443.51,128.44) .. (440.19,128.44) -- (422.13,128.44) .. controls (418.81,128.44) and (416.12,125.75) .. (416.12,122.42) -- (416.12,50.45) .. controls (416.12,47.13) and (418.81,44.44) .. (422.13,44.44) -- cycle ;
\draw  [fill={rgb, 255:red, 184; green, 233; blue, 134 }  ,fill opacity=1 ] (736.79,245.85) .. controls (740.11,245.85) and (742.81,248.54) .. (742.81,251.87) -- (742.81,340.03) .. controls (742.81,343.36) and (740.11,346.05) .. (736.79,346.05) -- (718.73,346.05) .. controls (715.41,346.05) and (712.72,343.36) .. (712.72,340.03) -- (712.72,251.87) .. controls (712.72,248.54) and (715.41,245.85) .. (718.73,245.85) -- cycle ;
\draw  [fill={rgb, 255:red, 184; green, 233; blue, 134 }  ,fill opacity=1 ] (328.14,294.37) .. controls (328.28,302.87) and (321.51,309.82) .. (313.01,309.89) -- (78.25,311.9) .. controls (69.74,311.97) and (62.73,305.14) .. (62.59,296.64) -- (61.8,250.46) .. controls (61.65,241.96) and (68.43,235.01) .. (76.93,234.94) -- (311.69,232.93) .. controls (320.19,232.86) and (327.2,239.69) .. (327.35,248.19) -- cycle ;
\draw  [fill={rgb, 255:red, 184; green, 233; blue, 134 }  ,fill opacity=1 ] (640.01,426.72) .. controls (640.16,435.22) and (633.38,442.17) .. (624.88,442.24) -- (316.12,444.88) .. controls (307.62,444.96) and (300.61,438.12) .. (300.46,429.62) -- (299.67,383.44) .. controls (299.53,374.94) and (306.3,367.99) .. (314.81,367.91) -- (623.57,365.27) .. controls (632.07,365.2) and (639.08,372.03) .. (639.22,380.54) -- cycle ;
\draw    (186.86,85.93) -- (219.25,85.93) ;
\draw [shift={(221.25,85.93)}, rotate = 180] [color={rgb, 255:red, 0; green, 0; blue, 0 }  ][line width=0.75]    (10.93,-3.29) .. controls (6.95,-1.4) and (3.31,-0.3) .. (0,0) .. controls (3.31,0.3) and (6.95,1.4) .. (10.93,3.29)   ;
\draw    (380.29,84.92) -- (412.68,84.92) ;
\draw [shift={(414.68,84.92)}, rotate = 180] [color={rgb, 255:red, 0; green, 0; blue, 0 }  ][line width=0.75]    (10.93,-3.29) .. controls (6.95,-1.4) and (3.31,-0.3) .. (0,0) .. controls (3.31,0.3) and (6.95,1.4) .. (10.93,3.29)   ;
\draw    (328.71,278.24) -- (412.91,234.62) ;
\draw [shift={(414.68,233.7)}, rotate = 152.62] [color={rgb, 255:red, 0; green, 0; blue, 0 }  ][line width=0.75]    (10.93,-3.29) .. controls (6.95,-1.4) and (3.31,-0.3) .. (0,0) .. controls (3.31,0.3) and (6.95,1.4) .. (10.93,3.29)   ;
\draw    (449.5,249) -- (521.93,199.41) ;
\draw [shift={(523.58,198.28)}, rotate = 145.6] [color={rgb, 255:red, 0; green, 0; blue, 0 }  ][line width=0.75]    (10.93,-3.29) .. controls (6.95,-1.4) and (3.31,-0.3) .. (0,0) .. controls (3.31,0.3) and (6.95,1.4) .. (10.93,3.29)   ;
\draw    (446.21,84.92) -- (521.85,128.46) ;
\draw [shift={(523.58,129.46)}, rotate = 209.92] [color={rgb, 255:red, 0; green, 0; blue, 0 }  ][line width=0.75]    (10.93,-3.29) .. controls (6.95,-1.4) and (3.31,-0.3) .. (0,0) .. controls (3.31,0.3) and (6.95,1.4) .. (10.93,3.29)   ;
\draw    (638.21,406.78) -- (710.1,308.19) ;
\draw [shift={(711.28,306.58)}, rotate = 126.1] [color={rgb, 255:red, 0; green, 0; blue, 0 }  ][line width=0.75]    (10.93,-3.29) .. controls (6.95,-1.4) and (3.31,-0.3) .. (0,0) .. controls (3.31,0.3) and (6.95,1.4) .. (10.93,3.29)   ;
\draw  [fill={rgb, 255:red, 224; green, 100; blue, 115 }  ,fill opacity=1 ] (736.79,114.27) .. controls (740.11,114.27) and (742.81,116.97) .. (742.81,120.29) -- (742.81,205.42) .. controls (742.81,208.74) and (740.11,211.44) .. (736.79,211.44) -- (718.73,211.44) .. controls (715.41,211.44) and (712.72,208.74) .. (712.72,205.42) -- (712.72,120.29) .. controls (712.72,116.97) and (715.41,114.27) .. (718.73,114.27) -- cycle ;
\draw    (631.04,165.89) -- (707.85,166.88) ;
\draw [shift={(709.85,166.9)}, rotate = 180.74] [color={rgb, 255:red, 0; green, 0; blue, 0 }  ][line width=0.75]    (10.93,-3.29) .. controls (6.95,-1.4) and (3.31,-0.3) .. (0,0) .. controls (3.31,0.3) and (6.95,1.4) .. (10.93,3.29)   ;
\draw    (742.81,160.83) -- (825.51,196.48) ;
\draw [shift={(827.34,197.27)}, rotate = 203.32] [color={rgb, 255:red, 0; green, 0; blue, 0 }  ][line width=0.75]    (10.93,-3.29) .. controls (6.95,-1.4) and (3.31,-0.3) .. (0,0) .. controls (3.31,0.3) and (6.95,1.4) .. (10.93,3.29)   ;
\draw    (744.24,300.5) -- (826.09,198.83) ;
\draw [shift={(827.34,197.27)}, rotate = 128.83] [color={rgb, 255:red, 0; green, 0; blue, 0 }  ][line width=0.75]    (10.93,-3.29) .. controls (6.95,-1.4) and (3.31,-0.3) .. (0,0) .. controls (3.31,0.3) and (6.95,1.4) .. (10.93,3.29)   ;

\draw (865.01,192.78) node [anchor=north west][inner sep=0.75pt]  [font=\huge]  {$D_{\phi }$};
\draw (172.48,36.89) node [anchor=north west][inner sep=0.75pt]  [font=\huge]  {$z$};
\draw (291.22,67.27) node [anchor=north west][inner sep=0.75pt]  [font=\huge]  {$G_{\theta }$};
\draw (128.11,242.97) node [anchor=north west][inner sep=0.75pt]  [font=\huge] [align=left] {Image dataset};
\draw (106.67,269.7) node [anchor=north west][inner sep=0.75pt]  [font=\huge]  {$\{x_{1} ,\ x_{2} ,\ \cdots ,x_{N}\}$};
\draw (365.43,378.14) node [anchor=north west][inner sep=0.75pt]  [font=\huge] [align=left] {Measurement dataset};
\draw (391.81,407.35) node [anchor=north west][inner sep=0.75pt]  [font=\huge]  {$\{y_{1} ,\ y_{2} ,\ \cdots ,y_{N}\}$};
\draw (423.39,10.59) node [anchor=north west][inner sep=0.75pt]  [font=\huge]  {$\kappa ^{g}$};
\draw (424.18,327.39) node [anchor=north west][inner sep=0.75pt]  [font=\huge]  {$x^{r}$};
\draw (560.51,146.2) node [anchor=north west][inner sep=0.75pt]  [font=\huge]  {$\mathcal{A}$};
\draw (725.07,355.73) node [anchor=north west][inner sep=0.75pt]  [font=\huge]  {$y^{r}$};
\draw (715.48,71.32) node [anchor=north west][inner sep=0.75pt]  [font=\huge]  {$y^{g}$};

\end{tikzpicture}}
\caption{Schematic block diagram illustrating our algorithm for unsupervised learning of the imaging system parameters. We learn a generative network $G_\theta$ that generates system parameters $\kappa^g$. Given a set of training images $\{x_1, x_2, \cdots, x_N\}$, we pass these images along with the parameters $\kappa^g$ through the imaging system to generate measurements $y^g$. The discriminative network $D_\phi$ is trained to distinguish between the generated measurements $y^g$ and actual measurements $y^r \sim\{y_1, y_2, \cdots, y_N\}$. Note that the image dataset and measurement dataset are \emph{unpaired}, independent, and disjoint of one another.}\label{fig:ambient}
\end{figure*}
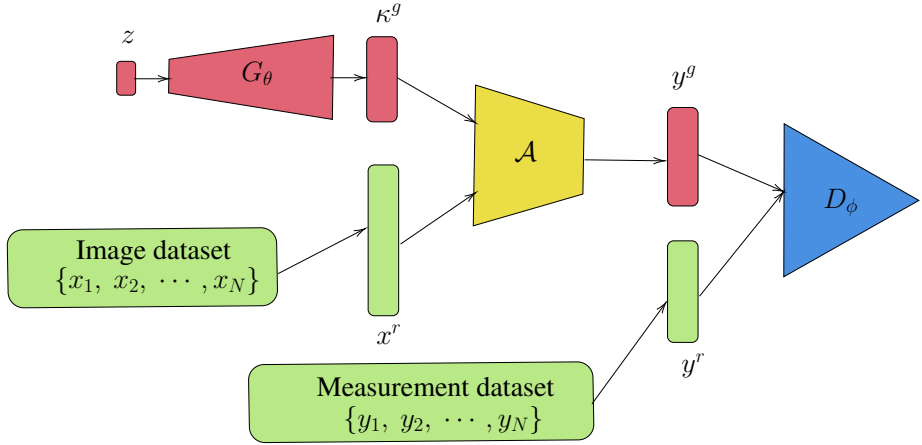

\paragraph{Blind Inverse Problems.} 
In many practical scenarios, the imaging system is not known exactly at the time of signal recovery, but can be modeled as belonging to a family of forward models $A_\kappa$, where $\kappa$ parameterizes the unknown components.
The uncertainty in the imaging system can stem from various sources, and can often be modeled statistically. Randomness may be introduced by the imaging hardware itself, such as fixed but unknown lens aberrations. Uncertainty may also arise from hardware-agnostic effects like camera shake \cite{Delbracio_2015_CVPR}, motion blur due to scene dynamics \cite{gong2017blur2mf}, or atmospheric turbulence \cite{mao2020imagereconstructionstaticdynamic}. In these cases, even if the imaging hardware is static, the overall system is not, necessitating estimation of the imaging system for each measurement capture.

Model-based algorithms for blind inverse problems jointly estimate both the clean image $x$ and the correct forward model $A_\kappa$ corresponding to the measurement data $y$ \cite{li2017bilinearinv,ling2018bilinearinv,krishnan2011blindeconv,Wipf2013RevisitingBB,ali2014blinddeconv}. This is typically accomplished by imposing priors on the image, $p(x)$, and the measurement model, $p(A)$ \cite{chung2023parallel}. 
While it is often feasible to learn $p(x)$ through generative modeling using available clean images $x \sim p(x)$, directly observing or learning $p(A)$ is more challenging. This is because the imaging system itself is not directly observed; rather, only the measurements $y \sim p(y)$ that follow Eq. \ref{eqn:forward_process}. Moreover, it is not practical to assume a training set of paired data of images and measurements, i.e. $\{(x_i, y_i)\}_{i\in [N]}$, which further complicates the learning of the imaging system.

\paragraph{Our approach.} 
We are interested in the setting where we do not have access to direct observations of the forward process or paired data. We assume there is some low-dimensional structure in the forward process, and we assume access to an independent training set of clean images, and an independent training set of measurements.
We build upon the principles of AmbientGAN \cite{bora2018ambientgan}, but with a crucial distinction: instead of learning the distribution of clean images, we learn the distribution of forward operators using the unpaired images and measurements as training data. Our underlying assumption is that the measurement process can be represented implicitly by a generative model due to some underlying low-dimensional structure.

This approach to learning the imaging system from unpaired examples naturally integrates with downstream system ID tasks that have model uncertainty (e.g. in blind deconvolution). Traditional blind deconvolution methods rely on alternating or joint estimation of the image and the blur kernel by assuming both are unknown and iteratively refining both components \cite{krishnan2011blindeconv,Wipf2013RevisitingBB,ali2014blinddeconv}. With our generative model-based framework, the learned distribution of forward operators $p(A)$ can be used to provide a robust prior for the measurement operator in these solvers. This prior can guide off-the-shelf blind deconvolution algorithms by constraining the possible forms of the blur kernel, improving convergence to accurate solutions even when no exact operator measurements are available \cite{chung2023parallel}. Our approach acts as a flexible plug-in for downstream tasks that require a statistical model for the imaging system.





\subsubsection*{Our Contributions}
\begin{enumerate}
    \item  We propose an unsupervised algorithm that learns a prior distribution over forward operators using only \emph{unpaired} clean images and corrupted measurements that are independent and disjoint from one-another. Forward operators are structured operators defined by a physical imaging system, and in this paper we consider Gaussian smoothing and motion blur, though other structure can also be incorporated.  We design a network architecture for the generative model which respects the physics of the imaging system and then leverage the loss function in AmbientGAN~\cite{bora2018ambientgan} to train the model.

    \item  We experimentally show that our method is robust to additive noise in the imaging system. 
    The experiments in Section~\ref{sec:noise_robust} show that our method is robust in low-SNR settings.

    \item Our learning algorithm is independent of the downstream system ID or signal recovery algorithm. In Section~\ref{sec:downstream_rest}, we show that this modular approach of learning system priors can be used in off-the-shelf state-of-the-art algorithms~\cite{chung2023parallel} for blind deconvolution. 

    \item In Section~\ref{sec:downstream_rest}, we evaluate our algorithm on the classical problems of Gaussian and motion deblurring. We find that training with imaging forward processes learned with our technique enables reconstruction quality close to that of models trained with direct access to imaging system forward process examples.
\end{enumerate}

These contributions provide a versatile and robust tool for enabling blind inverse imaging problem solvers, addressing limitations of previous models that rely on direct observations of measurement systems. 
While we focus on the application of blind deblurring here, our framework has the potential to be applied to more general structured blind inverse problems.

\section{Related Work}
\label{sec:related_work}

\subsection{Generative Modeling}
\label{subsec:gen_model}

The goal of generative modeling is to learn a probability distribution $p(x)$ with access to a training set of individual samples $\{x_i\}_{i=1,...N}$. This can be done by a variety of different ways \cite{kingma2022autoencodingvariationalbayes, goodfellow2014generativeadversarialnetworks,Kobyzev_2021,ho2020denoisingdiffusionprobabilisticmodels,song2021scorebasedgenerativemodelingstochastic}. Each approach has pros and cons \cite{Bond-Taylor2022}; here we focus on generative adversarial networks (GAN) and diffusion models as we use attractive properties of GANs for the task of learning the imaging model and we use diffusion models for the downstream task of blind inverse problems. 

GANs are a class of implicit generative models which learn to sample from the desired distribution $p(x)$ by solving a min-max optimization problem between competing generator $G_\theta$ and discriminator networks $D_\phi$. In their most basic form, GANs are trained by optimizing network weights $\theta, \phi$ using the optimization objective
\begin{align}
    \label{eqn:vanilla_GAN_train}
    \min_{G_{\theta}}\max_{D_{\phi}}  &\E_{x\sim p(x)}[\ln(D_\phi(x))] \nonumber\\ 
    &+ \E_{z\sim p_z(z)}[\ln(1-D_\phi (G_\theta(z)))].
\end{align}
There are many practical issues that arise when training GANs. For example, the inherent instability in solving a min-max problem can lead to mode collapse. To address these issues, a variety of alterations and improvements over the basic objective and training procedure above have been proposed such as Wasserstein GANs \cite{arjovsky2017wassersteingan} and discriminator augmentation \cite{karras2020traininggenerativeadversarialnetworks}. 

Diffusion models are a class of generative models that approximate a target data distribution \( p(x) \) by learning to reverse a process that gradually transforms data into noise over a continuous time interval \cite{ho2020denoisingdiffusionprobabilisticmodels, song2021scorebasedgenerativemodelingstochastic,karras2022elucidatingdesignspacediffusionbased}. They do this by learning to estimate the \emph{score function} \( \nabla_{x} \log p_t(x) \) of a family of intermediate distributions \( p_t(x) \), where \( p_t(x) \) is a progressively noised version of the original data distribution. These distributions \( p_t(x) \) are produced by adding noise to the data in a continuous or discrete fashion, controlled by a variance schedule.

The forward (or \emph{diffusion}) process that creates these noised distributions can be formalized as a \emph{Stochastic Differential Equation} (SDE), which models the continuous-time addition of noise:
\begin{equation}
\label{eqn:forward_sde}
    dx = f(x,t)dt + g(t)d\mathbf{\omega},
\end{equation}
where $f(x,t)$ is the drift term which controls the deterministic part of the transformation, $g(t)$ is the diffusion term that scales the noise added at each time step, and $\omega$ is Brownian motion. It is possible to learn the score of intermediate distributions in an unsupervised fashion using denoising score matching \cite{DSM2011}. With the score function in hand, samples can be drawn from $p(x)$ by solving the following SDE \cite{song2021scorebasedgenerativemodelingstochastic, karras2022elucidatingdesignspacediffusionbased}:
\begin{equation}
\label{eqn:reverse_sde}
    dx = [f(x,t)-g^2(t) \nabla_x\log p_t(x)]dt + g(t)d\bar{\omega},
\end{equation}
which can be solved using Euler integration and other higher order solvers.
A common assumption for training these models is direct access to clean samples
$\{x_i\}_{i=1,..., N} \sim p(x)$.

\subsection{Unsupervised Learning}
\label{subsec:us_learning}
Generative models are typically trained with access to a large dataset of signals from the desired distribution. In practice, however, this is not always possible. Specifically, since most signals of interest are collected in the natural world using physical measurement devices it is common that the underlying signal distribution is corrupted by the action of the measurement process (e.g., noise, low-resolution, etc.). A valid question then arises: when can we learn the signal distribution $p(x)$ with access only to measurements $y\sim p(y)$ where $y=A(x) + \eta$? 

Ultimately this requires varying alterations to the training procedure depending on the generative model type \cite{bora2018ambientgan, kelkar2023ambientflowinvertiblegenerativemodels, daras2023ambientdiffusionlearningclean,leong2023structurefromcorruption,kawar2024gsurebased}. To guarantee recovery even in the noiseless case, there must often also be assumptions over the underlying dimensionality of the $p(x)$ in addition to the structure of the measurement process $A$ used to collect data\cite{julian2023}. In the case that additive noise is present in the measurement data, recovery of $p(x)$ is often still theoretically possible \cite{julian2023}. The application of unsupervised learning in the context of inverse problems almost always assumes a known forward model $A$ and an unknown image distribution $p(x)$.

\subsection{Blind Inverse Problem Solvers with Deep Learning}
\label{subsec:bip_solvers}
As with traditional inverse problems, blind-inverse problems can be solved in a variety of ways. In the context of deep learning,  the first of such categories is end-to-end estimation. In this category it is common to assume a dataset of clean signal + corrupt measurement pairs $\{x_i, y_i\}_{i=1,...,N}$ which can then be used to directly train a neural network $f_\theta(\cdot): y\rightarrow x$ to invert the forward process $y=Ax + \eta$ (e.g., MMSE estimators), or training a time-conditional sampler to sample from the posterior distribution $p(x|y)$ \cite{Whang_2022_CVPR}. A key drawback to end-to-end techniques is that they struggle to adapt to distribution shifts in the corruption process at test time \cite{levac2022}. Additionally, those techniques which produce deterministic estimators have been shown to be provably sub-optimal when it comes to perceptual quality \cite{Blau_2018, ohayon2023reasonssuperioritystochasticestimators}. 

Another category leverages generative models as plug-and-play priors in a stochastic sampling scheme, for example using diffusion models\cite{levac2024,chung2023parallel, laroche2023fastdiffusionemdiffusion, bai2024blindinversionusinglatent}. These techniques typically assume independence and have separate priors for the image and forward operator. Perhaps the most popular of such approaches, Blind Diffusion Posterior Sampling \cite{chung2023parallel}, runs joint optimization through a parallel sampling procedure over the image and the measurement system:

\begin{multline}
\label{eqn:bdps_image}
    dx = \bigl( -\frac{\beta(t)}{2}x - \beta(t)\bigl[\nabla_{x_t}\log p(y|\hat{x}_0, 
    \hat{\kappa}) \\+ D_{\theta_x}(x_t,t)\bigr] \bigr)dt + \sqrt{\beta(t)}d\bar{\mathbf{\omega}}
\end{multline}
\begin{multline}
\label{eqn:bdps_kappa}
    d\kappa = \bigl( -\frac{\beta(t)}{2}\kappa - \beta(t)\bigl[\nabla_{\kappa_t}\log p(y|\hat{x}_0, 
    \hat{\kappa}) \\+ D_{\theta_\kappa}(\kappa_t,t)\bigr] \bigr)dt + \sqrt{\beta(t)}d\bar{\mathbf{\omega}} 
\end{multline}

Where $\beta(t) \in \mathbb{R}$ is a time dependent signal scaling, and $D_{\theta_x}$ and $D_{\theta_\kappa}$ are pre-trained diffusion networks on images and forward operators, respectively. This approach, and others like it, allow for more modular training and don't require paired data of clean images $x_i$ and measurement operators $A_\kappa$. They do, however, require access to direct observations of $A_\kappa$ which is not always possible. Recent approaches that do not assume distributional independence attempt to sample from the joint posterior \cite{Murata2023GibbsDDRM}.

\section{Our Approach -- Double Blind Imaging}
\label{sec:method}
As stated above, many SOTA approaches to solving blind inverse problems require information about the prior over possible measurement operators $p(A)$ \cite{chung2023parallel, laroche2023fastdiffusionemdiffusion}. We propose learning this from only unpaired samples of clean images and corrupted measurements. In other words, we use samples from the marginal distributions $x_i \sim p(x)$ and $y_j \sim p(y)$ respectively where it is assumed that 
\begin{equation}
    y=A_\kappa(x) + \eta
\end{equation}
where $A_\kappa \sim p(A_\kappa)$, and $\eta$ is additive noise of a known family (Gaussian with known variance in this work). It is also assumed that $A_\kappa$ and $x$ are independent random variables.

A schematic diagram of our approach is shown in Figure~\ref{fig:ambient}. Our method approximates $p(A_\kappa)$ implicitly by learning how to match the measurement distribution $p(y)$ by selecting an $x_i$ from the clean dataset and passing it through a synthetic measurement system sampled from some generator function $A_\kappa = G_\theta(z)$, where $z \sim \mathcal{N}(0,I)$. Importantly, we assume some structure about $A$ is known, e.g. through the parameterization $\kappa$. For example, for blind deconvolution, $\kappa$ defines a $k$-dimensional convolution kernel.

More specifically, we use the following GAN inspired objective:
\begin{align}
    \label{eqn:db_GAN}
    &\min_{G_{\theta}}\max_{D_{\phi}}  \E_{y\sim p(y)}\left[\ln(D_\phi(y))\right] \nonumber\\ 
    &+ \E_{z\sim p_z(z), x\sim p(x), \eta \sim p(\eta)}\left[\ln(1-D_\phi (G_\theta(z)x + \eta))\right].
\end{align}
Note here that the discriminator does not incentive the output of the generative process to create realistic images but rather it pushes to learn realistic measurements which may be corrupted images (e.g., blurred, noised, etc.). The objective above is similar to that of AmbientGAN \cite{bora2018ambientgan}, except that we learn the generator of imaging systems (degradations processes) instead of clean distributions of signals. 

Due to the implicit nature of GAN training, a natural solution to learning in the presence of additive noise can be achieved. This is done by simply adding the correct amount of random noise at training time to the generated measurements from the synthetic imaging system. This implicitly deconvolves the noisy measurement distribution we learn to match with the additive noise distribution used to corrupt our real measurement data distribution. 


\begin{figure*}[ht]
    \centering
    \includegraphics[width=0.99\linewidth]{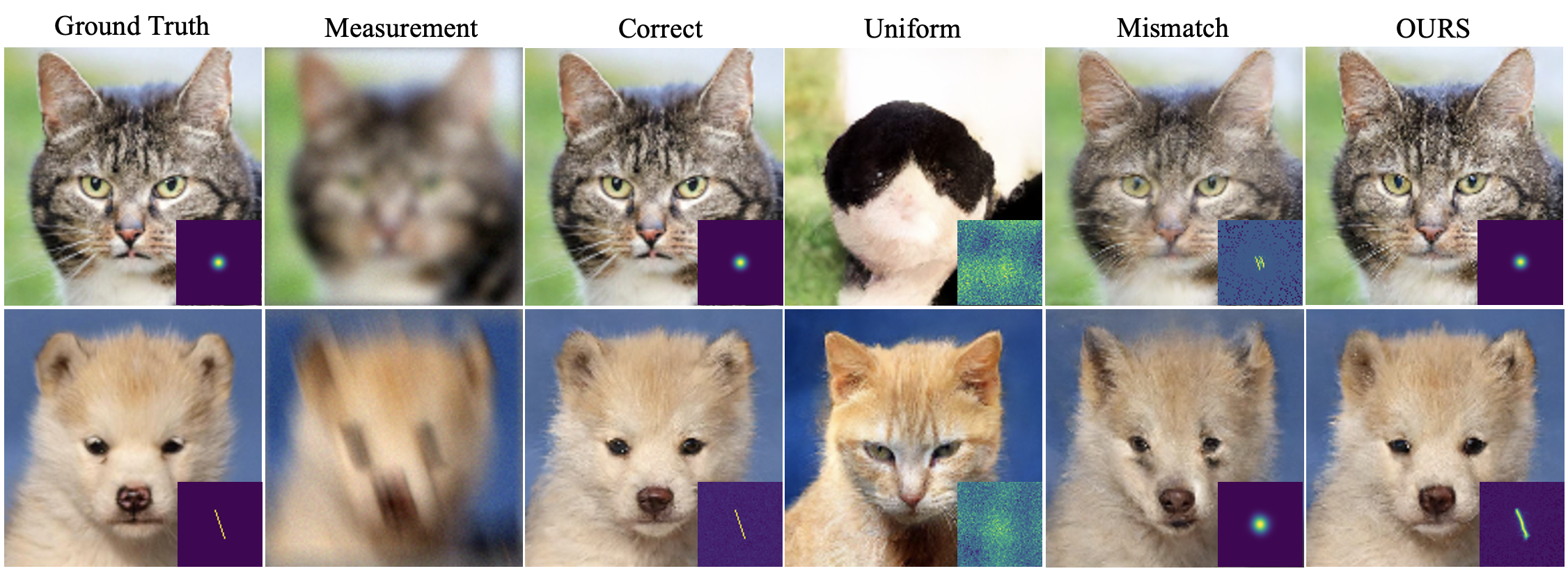}
    \caption{ BDPS reconstructions for blind Gaussian (top) and motion (bottom) deblurring on AFHQ $128\times 128$. Using various kernel priors.}
    \label{fig:BPDS_afhq_ex}
\end{figure*}
\section{Experiments}
\label{sec:experiments}
We focus our experiments on blind deconvolution, where $A_\kappa$ is a linear shift invariant system and $\kappa$ is a $k$-dimensional convolution kernel. While our approach can handle other low-dimensional structure, we leave more general forward operator constructions for future work.

\subsection{Point Spread Function Identification}
\label{sec:psf_id}
\begin{figure}[!t]
    \centering
    \includegraphics[width=0.99\linewidth]{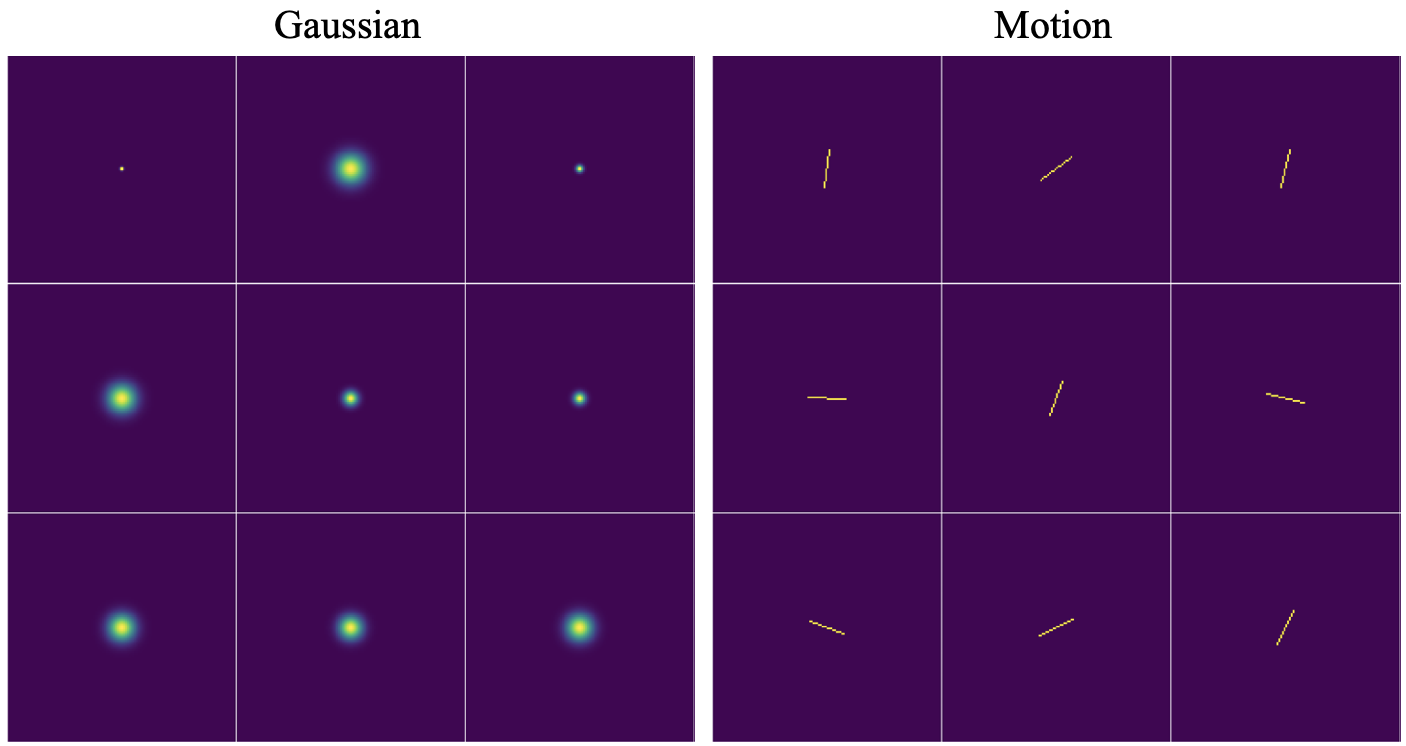}
    \caption{Ground truth point spread functions for both the Gaussian (left) and motion (right) examples.}
    \label{fig:gt_psfs}
\end{figure}

\begin{figure}[!t]
    \centering
    \includegraphics[width=0.99\linewidth]{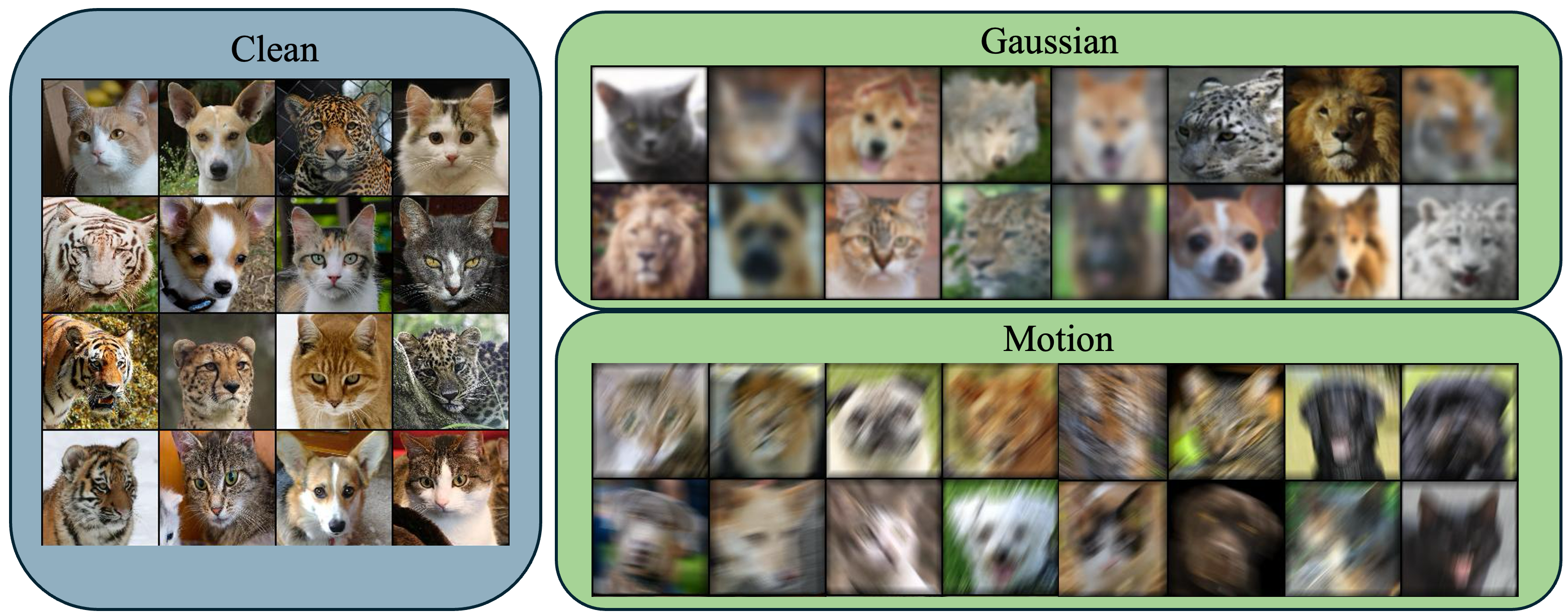}
    \caption{Example training split of unpaired data where the top row is for  Gaussian blur and the bottom row is motion blur.}
    \label{fig:train_split}
\end{figure}

We experiment with two different unknown point spread function (PSF) classes: unknown Gaussian blurring, and unknown motion blurring. In both cases, the kernel is assumed to have $k=128\times 128$ coefficients. \Cref{fig:gt_psfs} shows random PSF samples from each class. We conduct all training experiments on the AFHQ dataset \cite{choi2020starganv2diverseimage} resized to $128\times128$ images. We split the training dataset into two non-overlapping subsets of $N=7000$ images. One subset we keep as clean images $\{x_i\}_{i=1,...,N}\sim p(x)$ the other set of images is passed through random operators sampled from $\{A_{\kappa,j}\}_{j=1,...,N} \sim p(A_\kappa)$ to create a dataset $\{y_j\}_{j=1,...,N} \sim p(y)$. We emphasize that in this setup there are no image/measurement pairs shared between either subset of the training data. See Figure \ref{fig:train_split} for example of these subsets in the training dataset.

\subsection{Imaging Generator Architecture Choice}
\label{sec:arch_const}
\begin{figure}[!t]
    \centering
    \includegraphics[width=0.99\linewidth]{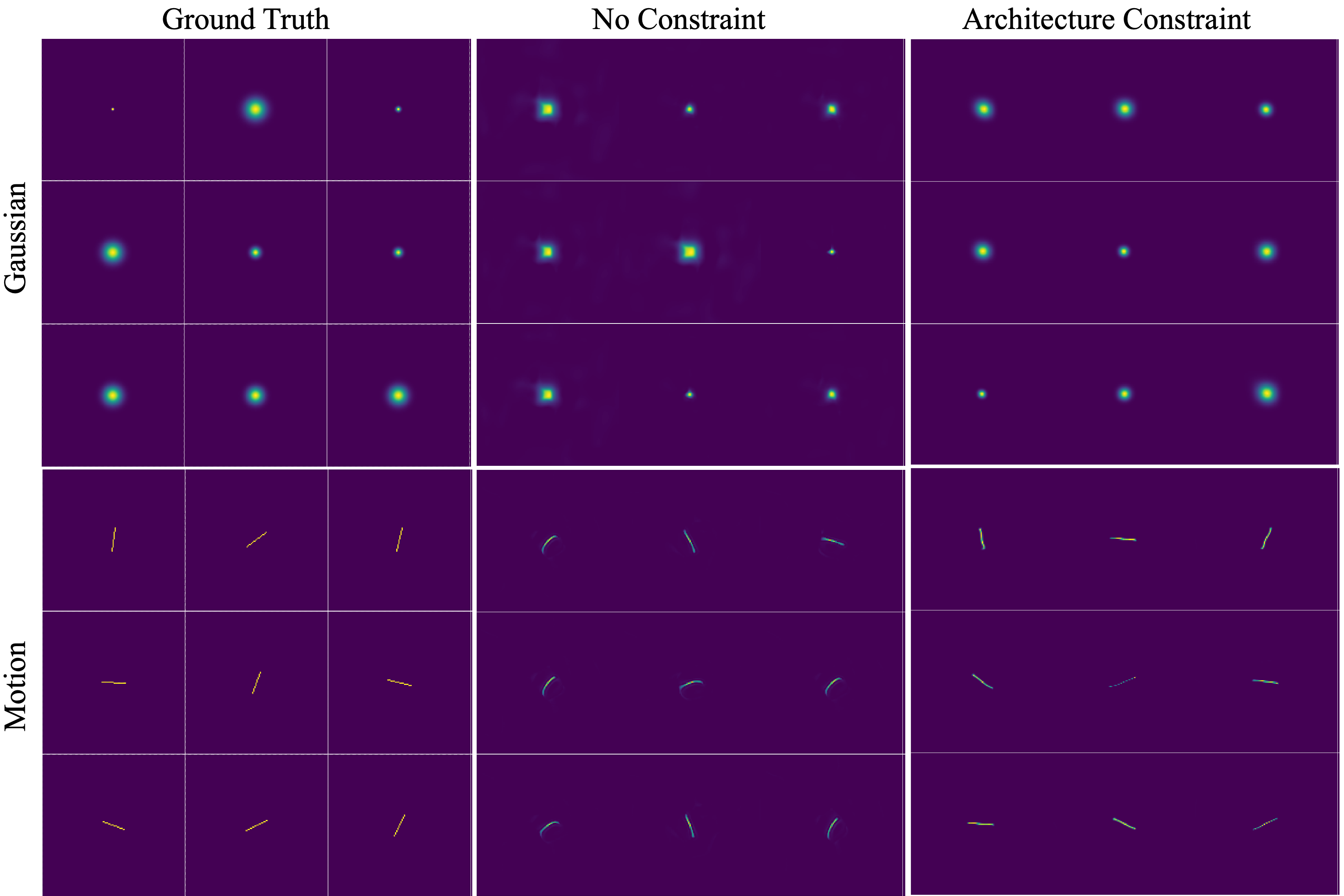}
    \caption{Comparisons of imaging systems learned with unconstrained vs. constrained generator architectures.}
    \label{fig:arch_const}
\end{figure}

In general, for linear inverse problems, the measurement model requires a generative architecture that outputs a matrix $A \in \mathbb{R}^{m\times n}$. If however, we know that the imaging system is inherently lower dimensional (e.g., linear shift invariant) we can adjust the architecture of our generative model accordingly. In the LSI case we consider here we have $k=128\times 128$ free parameters which must be learned. With this in mind we use the architecture proposed in \cite{karras2020traininggenerativeadversarialnetworks}. We used $17M$ parameters for the generator, and $30M$ for the discriminator. If we know certain things about the imaging system we can incorporate constraints directly into the imaging system generator itself. For example, we can add an energy constraint by including a softmax layer so as to only output normalized PSFs. To explore the effect of these architecture constraints we train two sets of models: \textbf{(1)} without softmax constraints \textbf{(2)} with softmax constraints. All models were trained for $22000$ steps using a batch size of $16$. We show that adding these architecture type constraints to the generative network can improve the representation quality of the imaging system. See \Cref{fig:arch_const} for example PSFs learned when using GAN architectures with and without normalization aware architectures. For the remainder of experiments we use a softmax output to normalize the PSFs.

\subsection{Noise Robustness}
\label{sec:noise_robust}
\begin{figure}[!t]
    \centering
    \includegraphics[width=0.99\linewidth]{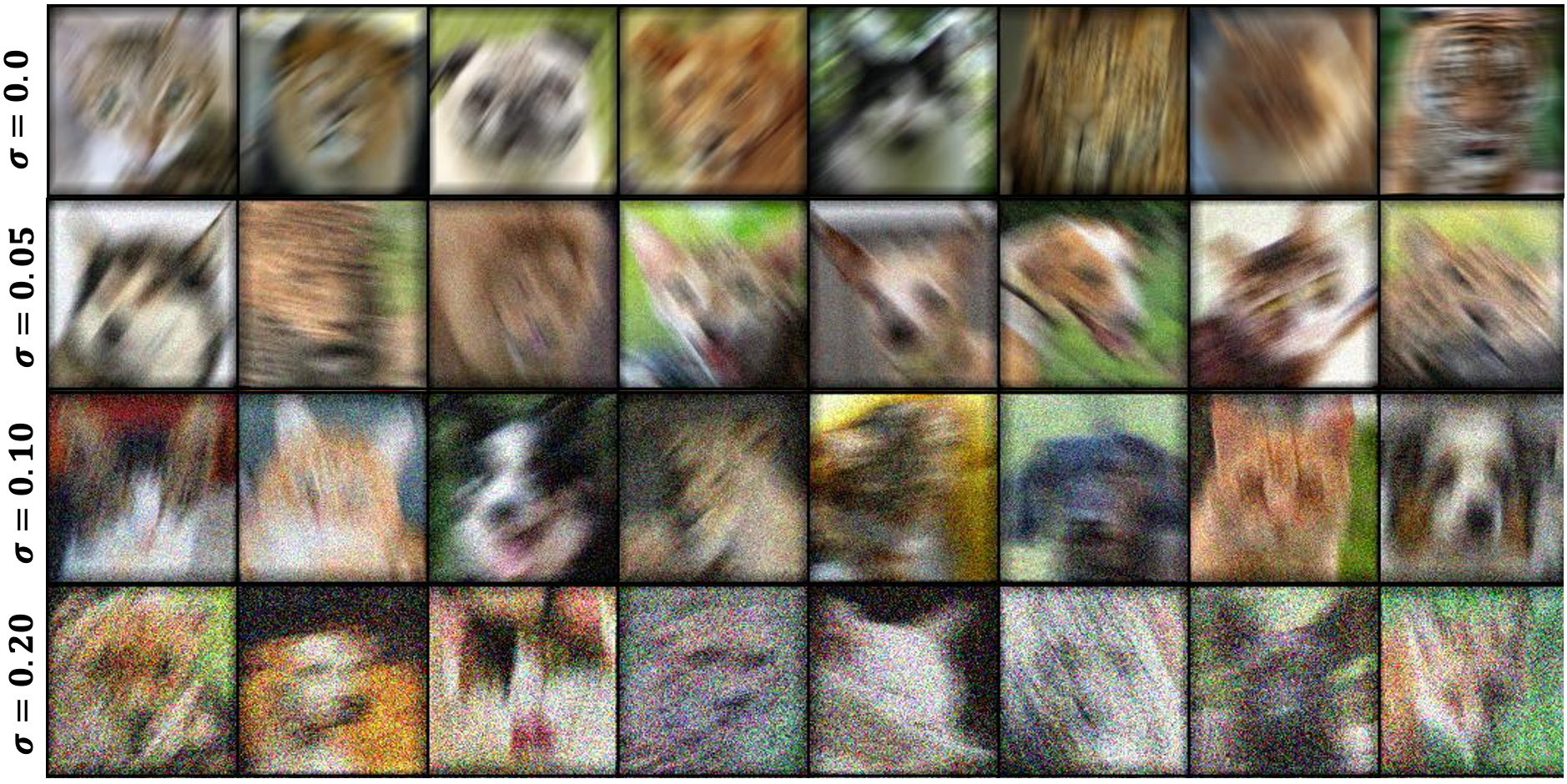}
    \caption{Noised measurement examples at various noise levels.}
    \label{fig:noised_meas}
\end{figure}

\begin{figure}[!t]
    \centering
    \includegraphics[width=0.99\linewidth]{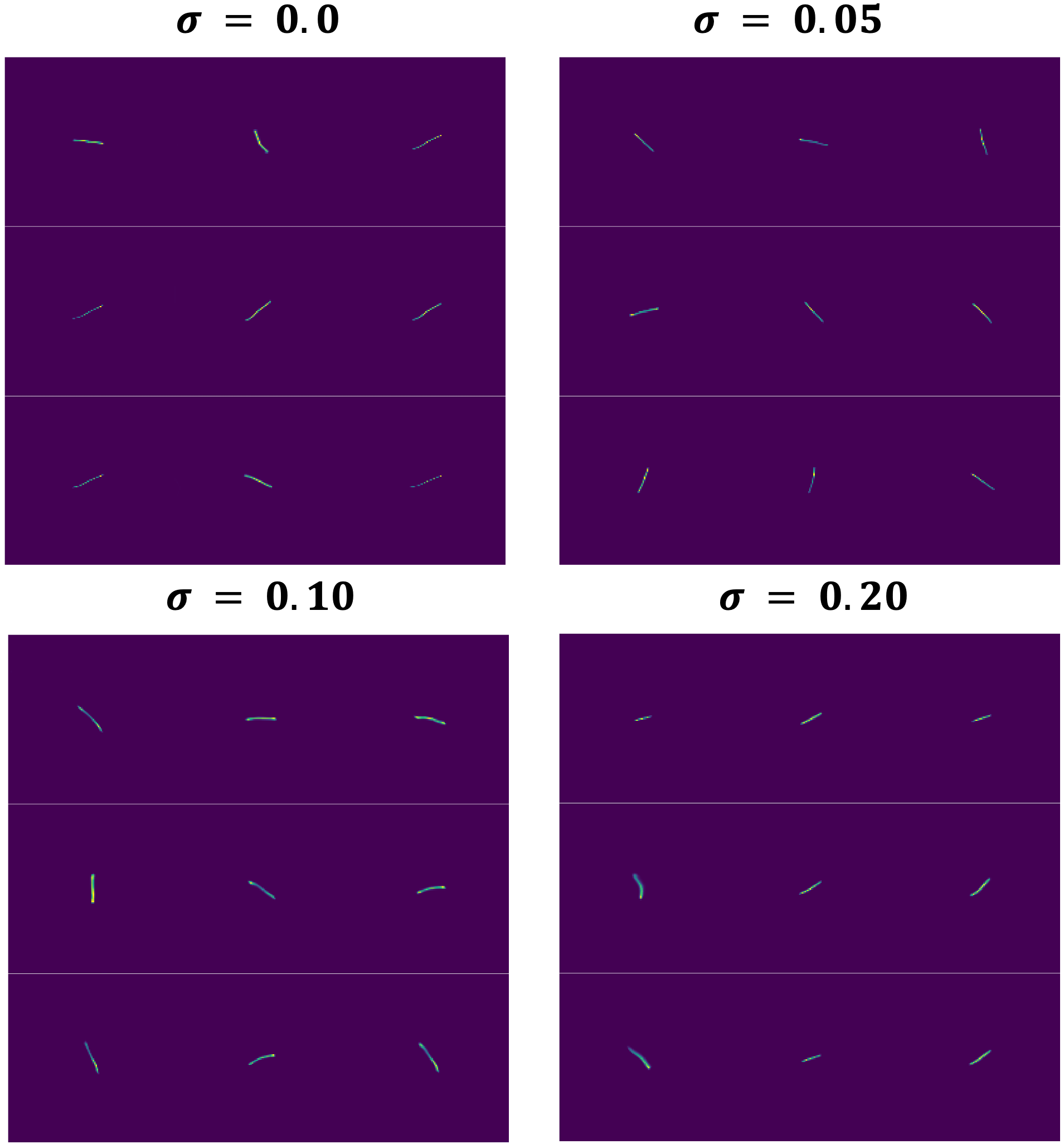}
    \caption{PSFs learned using the proposed technique on data at various noise levels.}
    \label{fig:noised_psf}
\end{figure}

To test our method's robustness to additive noise, we used the motion blur case and added varying amounts of Gaussian noise, $\sigma \in \{0.0, 0.05, 0.10, 0.20\}$, to the measurements. See \Cref{fig:noised_meas} for example measurements at each noise level. We applied our technique to learn the underlying distribution of PSFs in the presence of such noise with known value $\sigma$. We note here that noise was added a single time to all measurement data and that our networks did not see different noise instances for the same underlying blurred image at training time which ensures a more realistic training setting. See \Cref{fig:noised_psf} for random examples of learned motion PSFs at each noise level. 

\subsection{Double-blind diffusion posterior sampling}
\label{sec:downstream_rest}
Perhaps the most important measure of our proposed technique's utility is its use as a prior for solving blind inverse problems as a downstream task. There are many solvers that have been recently introduced \cite{chung2023parallel, laroche2023fastdiffusionemdiffusion,bai2024blindinversionusinglatent,Murata2023GibbsDDRM} which could have been used here. As our goal is not to propose a new blind inverse problem solver, but rather to show its use as a drop-in replacement, we choose a single popular method to evaluate our technique.

To test this we used the Blind Diffusion Posterior Sampling (BDPS) \cite{chung2023parallel} method which leverages parallel diffusion models: one trained on clean images, another trained on PSF kernels. Inference is then run using Eqs. \ref{eqn:bdps_image} and \ref{eqn:bdps_kappa}. We used both the AFHQ dataset at $128\times128$ and the FFHQ dataset at $256 \times 256$ for these experiments. For AFHQ, we trained a diffusion model over the $7000$ clean images in our training dataset. For FFHQ, we used the pretrained model from \cite{chung2023parallel}.

\begin{figure*}[!t]
    \centering
    \includegraphics[width=0.99\linewidth]{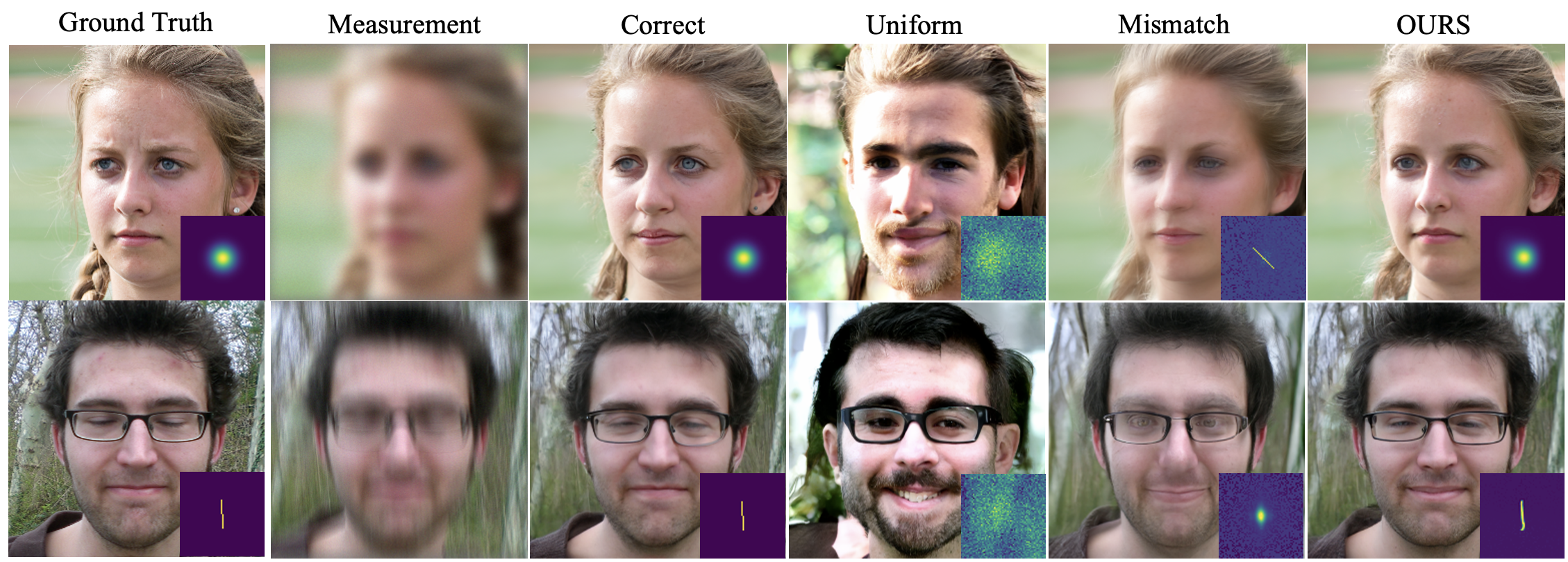}
    \caption{BDPS reconstructions for blind Gaussian (top) and motion (bottom) deblurring on FFHQ $256\times 256$. Using various kernel priors.}
    \label{fig:BPDS_ffhq_ex}
\end{figure*}

We then trained diffusion models over the PSFs for a variety of different cases, either using ground truth simulated kernels (i.e. first column of Figure~\ref{fig:arch_const}),  or kernels sampled from our trained generators (i.e. third column of Figure~\ref{fig:arch_const}). In both cases we used $7000$ PSF samples to train each diffusion model. The image and kernel networks were each trained for $150,000$ steps using a batch size of $32$. Inference was $1000$ steps using the repository provided by \cite{chung2023parallel}. For both the blind Gaussian and motion deblurring on AFHQ and FFHQ, we ran inference using a variety of kernel diffusion models. \textbf{(1)} The kernel model trained on ground-truth kernels (\textbf{Correct}); \textbf{(2)} a uniform prior over kernels (\textbf{Uniform}); \textbf{(3)} the incorrect kernel prior for the other task (e.g., Gaussian kernel model for motion deblurring and vice-versa)(\textbf{Mismatch}); and \textbf{(4)} the kernel trained on PSFs from our generative imaging technique (\textbf{Ours}). We display numerical results for both the blind Gaussian deblurring and motion deblurring cases on both the AFHQ and FFHQ datasets for both the image and kernel reconstructions in tables \ref{tab:afhq_image_metrics}, \ref{tab:ffhq_image_metrics} and \ref{tab:afhq_kernel_metrics},\ref{tab:ffhq_kernel_metrics} respectively. We show example reconstructions for the AFHQ and FFHQ experiments in Figures \ref{fig:BPDS_afhq_ex} and \ref{fig:BPDS_ffhq_ex} respectively.

\begin{table*}
  \centering
  \begin{tabular}{|c c c c c c c|}
    \toprule
    Kernel Prior & PSNR $(\uparrow)$ & SSIM ($\uparrow$) & LPIPS ($\downarrow$) & FID ($\downarrow$) & CMMD ($\downarrow$) & NIQE ($\downarrow$)\\
    \midrule
    Correct (Gaussian) & $\mathbf{22.17}$ & $\mathbf{0.563}$ & $\mathbf{0.1538}$ & $\mathbf{39.14}$ & $\mathbf{0.249}$ & $\underline{7.516}$  \\
    Uniform & $11.43$ & $0.207$ & $0.443$ & $111.82$ & $0.602$ & $9.828$ \\
    Mismatch (Motion) & $21.72$ & $\underline{0.558}$ & $0.203$ & $53.85$ & $0.587$ & $\mathbf{7.264}$  \\
    OURS & $\underline{21.81}$ & $0.550$ & $\underline{0.1595}$ & $\underline{40.00}$ & $\underline{0.280}$ & $7.608$ \\
    \midrule
    Correct (Motion) & $\mathbf{21.97}$ & $\mathbf{0.587}$ & $\mathbf{0.1401}$ & $\mathbf{40.76}$ & $\underline{0.345}$ & $\mathbf{6.520}$  \\
    Uniform & $11.63$ & $0.212$ & $0.4326$ & $102.57$ & $0.555$ & $9.773$ \\
    Mismatch (Gaussian) & $18.53$ & $0.393$ & $0.2608$ & $73.00$ & $0.892$ & $7.444$\\
    OURS & $\underline{21.43}$ & $\underline{0.555}$ & $\underline{0.1502}$ & $\underline{42.724}$ & $\mathbf{0.332}$ & $\underline{7.105}$ \\
    
    \bottomrule
  \end{tabular}
  \caption{Image reconstruction metrics on AFHQ $128\times128$ blind gaussian (top) and motion (bottom) deblurring using a variety of priors for the kernel.}
  \label{tab:afhq_image_metrics}
\end{table*}

\begin{table*}
  \centering
  \begin{tabular}{|c c c c c c c|}
    \toprule
    Kernel Prior & PSNR $(\uparrow)$ & SSIM ($\uparrow$) & LPIPS ($\downarrow$) & FID ($\downarrow$) & CMMD ($\downarrow$) & NIQE ($\downarrow$)\\
    \midrule
    Correct (Gaussian) & $\mathbf{25.53}$ & $\mathbf{0.721}$ & $\mathbf{0.1511}$ & $\mathbf{58.93}$ & $\mathbf{0.076}$ & $\underline{5.622}$  \\
    Uniform & $14.68$ & $0.363$ & $0.3863$ & $93.02$ & $1.048$ & $\mathbf{5.3886}$ \\
    Mismatch (Motion) & $24.835$ & $0.712$ & $0.2643$ & $86.00$ & $0.395$ & $8.0117$\\
    OURS & $\underline{25.16}$ & $\underline{0.714}$ & $\underline{0.1552}$ & $\underline{60.57}$ & $\underline{0.082}$ & $5.6688$ \\
    \midrule
    Correct (Motion) & $\mathbf{24.93}$ & $\mathbf{0.720}$ & $\underline{0.1602}$ & $\underline{63.27}$ & $\underline{0.092}$ & $5.923$  \\
    Uniform & $14.68$ & $0.362$ & $0.3893$ & $94.60$ & $0.990$ & $5.3212$ \\
    Mismatch (Gaussian) & $21.51$ & $0.597$ & $0.2455$ & $85.30$ & $0.629$ & $5.9278$\\
    OURS & $\underline{24.70}$ & $\underline{0.710}$ & $\mathbf{0.1539}$ & $\mathbf{60.36}$ & $\mathbf{0.090}$ & $5.7292$ \\
    
    \bottomrule
  \end{tabular}
  \caption{Image reconstruction metrics on FFHQ $256\times256$ blind gaussian (top) and motion (bottom) deblurring using a variety of priors for the kernel.}
  \label{tab:ffhq_image_metrics}
\end{table*}

\begin{table}
  \centering
  \begin{tabular}{|c c c c|}
    \toprule
    Kernel Prior & MSE $(\downarrow)$ & MAE ($\downarrow$) & MNC ($\uparrow$)\\
    \midrule
    Correct (Gaussian) & $\mathbf{0.0006}$ & $\mathbf{0.0072}$ & $\mathbf{0.9982}$   \\
    Uniform & $0.0162$ & $0.1100$ & $0.905$  \\
    Mismatch (Motion) & $0.0037$ & $0.0249$ & $0.9807$ \\
    OURS & $\underline{0.0009}$ & $\underline{0.0121}$ & $\underline{0.9960}$ \\
    \midrule
    Correct (Motion) & $\mathbf{0.0010}$ & $\mathbf{0.0048}$ & $\mathbf{0.9944}$  \\
    Uniform & $0.0173$ & $0.1116$ & $0.9023$ \\
    Mismatch (Gaussian) & $0.0032$ & $\underline{0.0109}$ & $0.9798$\\
    OURS & $\underline{0.0017}$ & $0.0144$ & $\underline{0.9898}$ \\
    
    \bottomrule
  \end{tabular}
  \caption{Kernel reconstruction metrics on AFHQ $128\times128$ blind gaussian (top) and motion (bottom) deblurring using a variety of priors for the kernel.}
  \label{tab:afhq_kernel_metrics}
\end{table}

\begin{table}
  \centering
  \begin{tabular}{|c c c c|}
    \toprule
    Kernel Prior & MSE $(\downarrow)$ & MAE ($\downarrow$) & MNC ($\uparrow$)\\
    \midrule
    Correct (Gaussian) & $\mathbf{0.0012}$ & $\mathbf{0.0098}$ & $\mathbf{0.9956}$   \\
    Uniform & $0.0140$ & $0.1019$ & $0.9239$  \\
    Mismatch (Motion) & $0.0065$ & $0.0503$ & $0.9733$ \\
    OURS & $\underline{0.0026}$ & $\underline{0.0260}$ & $\underline{0.9897}$ \\
    \midrule
    Correct (Motion) & $\mathbf{0.0017}$ & $\mathbf{0.0144}$ & $\mathbf{0.9939}$  \\
    Uniform & $0.0151$ & $0.1034$ & $0.9197$ \\
    Mismatch (Gaussian) & $0.0036$ & $\underline{0.0152}$ & $0.9775$\\
    OURS & $\underline{0.0034}$ & $0.0342$ & $\underline{0.9887}$ \\
    
    \bottomrule
  \end{tabular}
  \caption{Kernel reconstruction metrics on FFHQ $256\times256$ blind gaussian (top) and motion (bottom) deblurring using a variety of priors for the kernel.}
  \label{tab:ffhq_kernel_metrics}
\end{table}


\paragraph{Metrics.}
We report metrics over test sets of $N=200$ images for both AFHQ and FFHQ experiments. For images we report image-to-image based metrics such as SSIM \cite{SSIM}, PSNR, and LPIPS as well as reference free metrics such as FID, CMMD \cite{CMMD}, and NIQE \cite{NIQE}. To evaluate the quality of the reconstructed kernel we use MSE, MAE, and MNC \cite{MNC}.

\section{Results and Discussion}
\label{sec:discussion}
From our experiments we see that for both Gaussian and motion blurring, we are able to recover useful information about the underlying imaging system. 
Specifically, we can implicitly represent the imaging system prior using a generative model trained on unpaired sets of images and corrupted measurements. Samples from the generative model closely match true samples drawn from the respective distribution, even in the presence of noise (Figure~\ref{fig:noised_psf}).

Successful recovery of the PSFs is feasible because our generative model and forward model take advantage of a prior structure. In the case of the forward model, we assume in this work linear shift invariance, which implies the forward operator as parameterized by at most $n$ degrees of freedom. In practice, following \cite{chung2023parallel}, we further constrain this to $k<n$ to model localized blurring. When creating a ``fake'' sample $y$ to compare to a true sample $y_j$, we use the explicit knowledge that the measurement operator represents a convolution. Other structure could be imposed, for example to represent nonlinearities such as phase retrieval \cite{leong2023structurefromcorruption}, turbulence \cite{chan2022turbulence}, or other degradation such as quantization, saturation, and gamma correction \cite{Anger2018ModelingRD}.

As a simple example, in Figure~\ref{fig:arch_const} we showed that by modifying the architecture of the generator representing the imaging system we can inject domain knowledge about the physics of our imaging system to guide the learning procedure to more realistic imaging systems. 
Specifically, we observed that kernels produced from the softmax-constrained architecture provided more isotropic kernels for Gaussian and straighter kernels for motion than did their respective unconstrained generator counterparts. This is very important, in that it allows an explicit and relatively straight forward process for including knowledge of imaging physics to guide the learning procedure. Although not explored further in this paper, it is possible to enforce other useful properties in network architectures such as equivariance \cite{herbreteau2024NormEqNN, Chen_2022_CVPR} which have proven to be very useful in other signal recovery settings in imaging \cite{Chen_2022_CVPR, julian2023}.

Additionally, we displayed the noise robust properties of our technique which requires only knowing the statistics of the additive noise distribution (i.e., mean/variance of Gaussian noise). Even at very high noise levels $\sigma = 0.20$ (\cref{fig:noised_meas}) we were able to recover meaningful kernel representations (\cref{fig:noised_psf}). This is in part owed to the implicit distribution deconvolution properties of adversarial training which can be achieved by simply adding the correct additive noise at training time. We note that it is not unrealistic to know the additive noise statistics of a given imaging measurement process through classical estimation \cite{noise_var2006} or to even correct for it with access only to corrupted measurements \cite{metzler2020unsupervisedlearningsteinsunbiased, tachella2024unsureunknownnoiselevel}. For example, in MRI this is done with a simple pre-scan \cite{Kellman2005}.

Our method is useful for downstream tasks that require access to a prior model of the measurement system without direct access to imaging system examples at training time. A straightforward application of this is solving blind inverse problems. While a variety of methods could be used, including end-to-end learning, here we focus on BDPS because our prior is a straightforward drop-in replacement. We emphasize that our technique is not meant to compete with these methods, but rather provide the ability to do ``double-blind'' imaging.
With this in mind, we observed that by using our technique to learn the distribution of imaging systems we were able to provide both numerical (Tables \ref{tab:afhq_image_metrics}, \ref{tab:ffhq_image_metrics},
\ref{tab:afhq_kernel_metrics}, and \ref{tab:ffhq_kernel_metrics} ) and qualitative results (Figures \ref{fig:BPDS_afhq_ex} and \ref{fig:BPDS_ffhq_ex}) which demonstrate that the prior we learn over imaging systems only gives minimal degradations in restoration performance when compared to using true examples of imaging systems to train on.


A key drawback of our approach, however, is that to use our technique with diffusion based solvers required a two-stage training procedure: \textbf{(1)} training a GAN according to Section \ref{sec:method}, followed by \textbf{(2)} training a diffusion model on samples from the GAN in the previous step. As stated previously, there are unsupervised methods for training diffusion models \cite{daras2023ambientdiffusionlearningclean}. However, they explicitly require paired access to the measurement model and measurements. This was the primary motivator for using a GAN style training objective for our method rather than newer SOTA supervised generative techniques such as diffusion. Other recent approaches have also explored this paradigm \cite{leong2023structurefromcorruption}.
\section{Conclusion}
\label{sec:conclusion}
We proposed and validated a generative technique to learn stochastic imaging systems from unpaired + noisy data to assist in downstream imaging tasks such as solving blind inverse problems.
The training scheme allows for easy incorporation of prior physical system knowledge through architecture construction and can easily handle noisy data by adding appropriately matched noise at training time.
We demonstrated our technique for recovering the distributions of two blind deconvolution tasks: Gaussian blur and motion blur.
Finally, we demonstrated that the measurement system distribution which we learned can be very helpful in downstream blind inverse problem tasks such as image recovery. There is still a great deal of work which must be explored to \textbf{(1)} remove the need for two step training to use SOTA blind solvers \textbf{(2)} evaluate performance on more general linear and nonlinear imaging systems.
\section{Acknowledgments}
\label{sec:Acknowledgments}
This work was supported by NSF IFML 2019844, NSF CCF-2239687 (CAREER), Google Research Scholars Program.

{
    \small
    \bibliographystyle{ieeenat_fullname}
    \bibliography{main}
}


\end{document}